# An Edge-Based Resource Allocation Optimization for the Internet of Medical Things (IoMT)

Eyhab Al-Masri

School of Engineering and Technology, University of Washington Tacoma
*Corresponding Author: Eyhab Al-Masri, Email:ealmasri@uw.edu

**Abstract**

As the number of Internet of Medical Things (IoMT) increases, the need for performing on-premises tasks within hospitals or medical centers also increases. Many healthcare organizations are progressively embracing or adopting an edge computing paradigm such that computationally intensive tasks can be processed at the edge of the network in order to avoid latency and network reliability issues associated with offloading tasks to the cloud. The problem, however, hospitals or medical centers may not be equipped with sufficient computing resources that can process advanced ML or AI tasks efficiently. In addition, some tasks may not be easily offloadable or contain sensitive patient healthcare data which increases the risks of having malicious attacks. In this paper, we extend our Edgify resource provisioning framework to consider the task offloading of healthcare applications' involving patients' data as a multiple criteria decision making (MCDM) process that often involves multiple conflicting criteria (e.g. data privacy risks, costs, latency, network reliability, among others). We evaluate our proposed framework through a number of experiments which demonstrate the usefulness and effectiveness of employing our optimization approach within hospitals or medical centers.

## Introduction

The Internet of Thing (IoT) has brought about new opportunities and challenges in the deployment of healthcare applications. A considerable progress has been achieved over the past few years in building powerful healthcare applications with advanced Artificial Intelligence (AI) capabilities [1]. In addition, advancements in sensing and actuating technologies contributed to the formation of the Internet of Medical Things (IoMT), an emerging paradigm that provides foundational infrastructure for connecting medical devices, applications and healthcare services.

IoMT aims to improve the delivery of healthcare services by integrating healthcare data, processes and medical devices or mobile healthcare applications [2, 3, 4]. IoMT evolved due to the rapidly increasing number of medical devices and systems that are capable of capturing and transmitting medical data. As a result, IoMT is likely to play an increasingly important role in improving the operational productivity of healthcare organizations while improving the speed and accuracy of medical diagnosis and treatments in the upcoming years [5].

However, there are major research challenges associated with the capturing and processing of medical data for building effective health-related information workflows. Medical devices may have limited computational capabilities for processing advanced AI or ML tasks. Therefore, medical devices may require offloading advanced tasks that transcend beyond their basic hardware capabilities to more powerful computational resources in order to maintain the required speed and accuracy of medical diagnosis or treatments and patient monitoring in real-time.

Offloading IoMT tasks may often result in service performance degradation or be associated with network security threats or vulnerabilities involving medical data (e.g. patient and performance data). In addition, IoMT tasks travelling through the network is subject to latency issues which may result in longer response times for completing the tasks. What is therefore desirable is to determine the offloadability of tasks running on IoMT systems within hospitals.

The offloadability of a task is defined as an IoT system's ability to determine whether a task can be offloaded to one or more resources within a similar or different network [6]. To this extent, it is critical for IoMT systems within hospitals to determine which tasks can be offloaded and ones that need to be processed within a local environment. The problem, however, identifying the offloadability of medical tasks often involves multiple conflicting criteria. Hence, we need to treat task offloadability in IoMT systems as a multi-criteria decision making process that often involves a number of decision variables having various constraints.

Many healthcare applications within hospitals often operate within premises or through a local area network [11]. Some applications may require processing certain tasks outside the scope of the local network (e.g. cloud). This makes the problem of provisioning resources within a hospital environment for completing advanced operations very challenging. This problem can often be solved using edge-computing, a paradigm for shifting computational resources closer to end-user devices making it an ideal solution for completing advanced tasks within hospital environments. Unlike the cloud, however, edge computing environments often have limited computational resources [6].

Efficiently provisioning computational resources across edge-computing environments becomes inevitable. In this paper, we extend our Edgify framework for supporting IoMT environments. Our Edgify framework can be employed across hospital environments for allocating computational more effectively. Edgify can be integrated within the IT infrastructure of hospitals or reside in a micro datacenter (MDC) operating nearby hospitals. Due to the increasing demand of operating healthcare-related operations at the edge of the network, micro datacenters are likely to play an increasingly vital role in maintaining healthcare applications and services to meet their demands while effectively meeting response time guarantees.

To address the challenges with having limited resources across edge-based hospital environments, we introduce an optimization strategy that considers the allocation of resources as a multi-criteria decision making problem. By optimizing the resources required for completing computational tasks, it is then possible to effectively decentralize the data processing across hospital environments more efficiently and determine the offloadability of tasks to external resources (e.g. cloud).

The rest of this paper is organized as follows. Section II introduces the architecture of our proposed IoMT-based Edgify framework [5]. The experimental results and evaluation of our proposed optimization strategy for edge-based resource allocation are discussed in Section III. Finally, the conclusion and future work are provided in Section IV.

## Edgify: An Edge-based IoMT Framework

The proliferation of IoMT devices at the edge of networks requires efficient techniques for decentralizing data processing tasks or workloads involving patient data. To avoid the shortcomings with sending patient or performance data to cloud resources which are likely to be located far away from the deployed location of IoMT devices, it would be ideal to process tasks within a hospital environment assuming it is equipped with the necessary edge nodes for performing advanced computing operations. Figure 1 presents the Edgify framework for supporting IoMT environments.

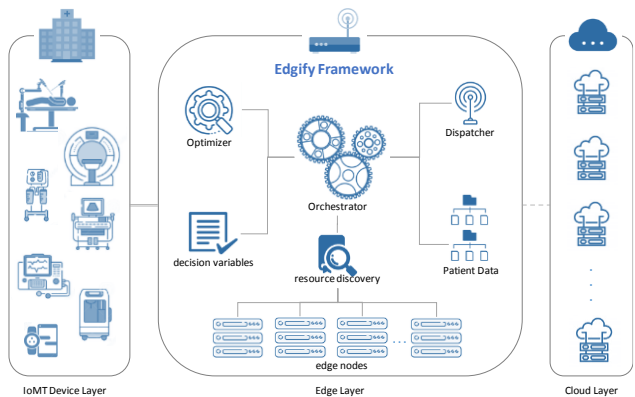

Fig. 1. Edgify Framework for Internet of Medical Things (IoMT).

As presented in Figure 1, the Edgify framework resides at the edge layer. This layer can be located within or in close proximity to a hospital environment. IoMT devices such as surgical robots, point of care (POC) devices, clinical wearables, medical sensors, smart pills, among many others. There are many advantages for processing workloads or tasks within an edge layer most notably reducing the time taken for conducting medical diagnosis and treatment, faster patient data access and improving the accessibility of patient and performance data, which can be used for processing insurance claims.

Because IoMT brings together a number of technologies, processing tasks may require applications that process very large volumes of real-time data collected through IoMT devices. To expedite the process of intelligently discovering insights within IoMT collected data, it is imperative to optimize the process of allocating resources within a hospital environment. Tasks or operations that require additional resources beyond those that exist within an edge environment can then be offloaded to the cloud for further processing. Therefore, it is essential to determine which tasks can run within the boundaries of the local environment and ones that can be offloaded to the cloud.

Edgify consists of a number of components that improves the process of offloadability of tasks. This is achieved through the optimization of edge-based resources in a decentralized manner prior to any offloading decisions to the cloud. Hence, the Edgify framework contains an orchestrator component that control and manages existing resources within the edge layer. An orchestrator creates a matrix, which is used by the optimizer component for determining an optimal resource allocation plan within the hospital environment.

Given that offloading requests often involve multiple conflicting criteria, Edgify employs multiple-criteria decision analysis (MCDA) methods such the Technique for Order Preference by Similarity to Ideal Solution (TOPSIS). The decision variables component is used to setup the optimization process and enables administrators to identify the importance level associated with each decision variable. To this extent, Edgify supports a number of predefined templates that can serve multiple types of IoMT applications. For example, latency-sensitive applications can be associated with a mission-critical predefined template ensuring that tasks in this category are given higher priority over non-mission critical tasks. We identify four main templates or IoMT workload types including: (a) cpu-centric, (b) memory-centric, (c) compute-centric, and (d) cost-centric.

The resource discovery component is used to discover existing resources within the edge layer. Resources at the edge of the network can vary in terms of the magnitude of hardware capabilities such as processing unit, memory, disk read/write rates, network receive/send rates, among others. The dispatcher component identifies the cloud-based resources where offloadable tasks can be dispatched.

## Evaluation and Results

For evaluating our Edgify IoMT optimization strategy, we considered the GWA-T-12 Bitbrains dataset [8]. The dataset contains metrics of 1,250 virtual machines (VMs) from a distributed datacenter. The dataset provides tracing data which are relevant for mapping edge-based resources such as (a) processor capacity provisioned, (b) processor usage in MHz, (c) memory capacity provisioned (KB), (d) memory usage (KB), (e) disk read/write rates (KB/s) and (f) network send/receive rates (KB/s). In addition, we used these attributes to derive two more attributes represent both cost and energy consumption models.

For the cost attribute, we employed Oracle's IT Infrastructure Power Calculator [9] and used the processor type as Intel Xeon 8260 2.4GHz, 165W, 24 Core and mapped the configuration items to those of the GWA-T-12 Bitbrains attributes. Using Oracle's calculator, we compute the total power of the system in Watts. We use this information to derive a cost model used for computing a cost estimation of the energy use relevant to the consumption of the required edge-based resources [10].

By using the energy usage and cost attributes, it is then possible to include these as part of the multiple-criteria decision analysis process. Using the defined decision variables, we use the Technique for Order Preference by Similarity to Ideal Solution (TOPSIS) decision method to identify the relevant resources available for completing an edge-based task or operation. In cases resources do not exist at the edge environment, Edgify then can offload tasks to other decentralized resources either through another edge layer within close proximity of the hospital (e.g. nearby datacenter) or to the cloud.

For evaluating the task computation offloading strategy, we primarily focus on the efficient and dynamic allocation of edge-based resources. That is, by efficiently allocating resources within the edge of the network, it is then possible to optimize the performance of IoMT applications and therefore enhance the offloading process. Because tasks can vary in terms of the requirements of hardware resources that need to be allocated, we identify a number of categories representing the task or operation type. Table I presents the various edge-based IoMT operation types that may exist within a hospital environment.

TABLE I
Edge-based IoMT Operation Types

| Task Type | Description |
|---|---|
| compute-centric 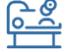 | a mission-critical and time-sensitive task that requires extensive data processing (e.g. cancerous diagnosis requiring advanced AI or DL processing) |
| cost-centric 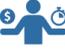 | a non-mission critical but somewhat time-sensitive task that reasonably requires efficient processing in terms of CPU, memory and most importantly cost (e.g. identifying possible treatment for specific medical case) |
| cpu-centric 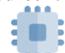 | similar to that of compute-centric except that cost is not a major factor for making decisions and reasonably requires sufficient memory for executing the task |
| memory-centric 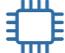 | a tasks that involves processing of incoming data through a network medium or one that requires processing an advanced data structure involving volatile memory |
| general 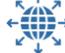 | a task that does not need specific preference in terms of CPU, memory, disk I/O, connectivity, cost or energy; a type of task that is general and all decision variables are treated of equal importance |

The type represents the priority associated with specific attributes and we assume that an application can provide the task or operation type when requesting edge-based resources to be allocated through Edgify's orchestrator component shown in Figure 1. Based on the task types, we evaluated our Edgify IoMT resource allocation strategy based on a number of use cases which are described below.

*A. Compute-Centric & Cost-Centric Use Cases*

A compute-centric operation is a mission-critical and time-sensitive task. In the context of IoMT, mission-critical tasks involve the use of a system or device that is considered essential for the hospital and/or patient. The continuing operation of such task is indispensable and any failure or disruption can be severe on the hospital operations or when operating on patients. An example of a compute-centric operation is one that is conducted by an active surgical robot that captures images for further processing and needs to provide recommendations to a surgical team for making accurate decisions involving a patient while in an operating room. Cost and energy consumption is not of the same level of importance as CPU and memory when executing a compute-centric task.

The cost-centric is somewhat similar to that of compute-centric but cost and energy consumption play a more significant role in the decision-making. The significance or priority is associated as weights to the TOPSIS decision method. We conducted an experiment utilizing the GWA-T-12 Bitbrains dataset where TOPSIS generates an optimization score (or performance score) which we refer to as Edgify Score.

We assume that the 1250 VMs are distributed throughout an edge-based hospital environment and Edgify encounters a task requesting resources to be allocated. Edgify uses TOPSIS to identify suitable resources in a dynamic manner and makes a decision of the best or most ideal resource that can be employed for executing the task. To this extent, we conducted an experiment for requesting resources for a compute-centric and cost-centric task types. We use the same edge-based nodes (or resources) for comparison purposes in terms of the overall performance score (Edgify Score). We also include the general case for comparison purposes. Results from this experiment are shown in Figure 2.

As presented in Figure 2, the yellow bars represent the performance scores of ten edge nodes ($e_1$ to $e_{10}$). Each edge node very in terms of the hardware capabilities it possesses. Depending on the task requirements, Edgify identifies the available resources that can satisfy these requirements. That is, we use Edgify to recommend the best available resource that can be allocated for executing or processing a task. In case of a compute-centric task, Edgify identifies that edge node 6 (or $e_6$) is the most suitable resource for executing this task. Similarly, the cost-centric use case identifies that edge node 6 (or $e_6$) is also suitable for executing the same task. Table II presents a cross-section of the normalized data values associated with the edge nodes (or resources) which are based on the GWA-T-12 Bitbrains dataset.

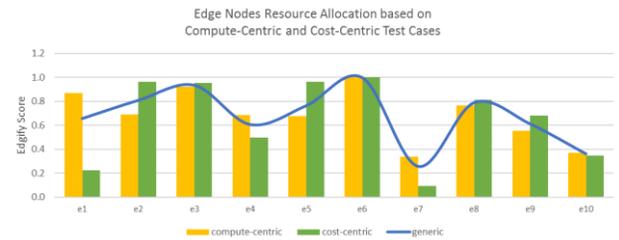

Fig. 2. Results from Compute- and Cost-Centric Task Allocation Experiment

For the compute-centric template, we employed a priority rate of 45% to CPU usage and available memory and 10% to cost and power, respectively. As can be seen in Table II, edge node 6 (or $e_6$) has the highest Edgify optimization score followed by edge node 1 (or $e_1$). Edge node 6 (or $e_6$) is among edge resources that have low CPU usage and available memory. Therefore, Edgify optimization strategy recommends a well-balanced solution in terms of both CPU and memory. In addition edge node 1 (or $e_1$) is ranked third in that list by Edgify since it has the highest memory availability. However, the resource has 52% CPU availability (e.g. somewhat busy). As shown in Figure 1, edge node 1 (or $e_1$) is ranked third by the Edgify framework for the compute-centric use case but ranks it ninth in terms of the cost-centric since it is clearly the second highest most expensive resource to be employed (e.g. 0.50 for cost attribute).

TABLE II
Edge Nodes Resource Capacities (Compute-Centric Use Case)

| Edge Node | CPU Usage | Available Memory | Cost | Power | Edgify Score |
|---|---|---|---|---|---|
| $e_1$ | 0.48 | 0.63 | 0.50 | 0.41 | 0.87 |
| $e_2$ | 0.18 | 0.18 | 0.07 | 0.16 | 0.69 |
| $e_3$ | 0.15 | 0.34 | 0.09 | 0.23 | 0.92 |
| $e_4$ | 0.37 | 0.34 | 0.32 | 0.35 | 0.68 |
| $e_5$ | 0.07 | 0.05 | 0.03 | 0.20 | 0.68 |
| **$e_6$** | **0.13** | **0.38** | **0.07** | **0.22** | **1.00** |
| $e_7$ | 0.50 | 0.20 | 0.54 | 0.43 | 0.34 |
| $e_8$ | 0.23 | 0.29 | 0.17 | 0.28 | 0.77 |
| $e_9$ | 0.29 | 0.19 | 0.23 | 0.31 | 0.55 |
| $e_{10}$ | 0.42 | 0.17 | 0.39 | 0.37 | 0.37 |

*B. CPU-Centric, Memory-Centric, Cost & Energy Use Cases*

We evaluated our Edgify task allocation strategy using more specialized use cases that focus primarily on a central factor or attribute. For example, the CPU-centric task type focuses on the CPU usage as the primary attribute having minimal or no weight associated to the other remaining attributes. For this experiment, we compare the outcomes of the CPU-centric and memory-centric to that of cost, energy and generic templates. Results from these experiments is shown in Figure 3.

As part of this experiment, we considered the same subset of edge nodes as in the generalized use cases for illustrating the effectiveness of the Edgify optimization. As presented in Figure 3, edge node 1 (or $e_1$) has the largest stacked segment for memory-centric (blue) which corresponds with the data provided in Table II having 0.63 (or highest) available memory compared to the rest of the encountered edge nodes.

For considering the CPU-centric, Edgify identifies edge node 5 (or $e_5$) as being the optimal resource (green segment). This result aligns with the data provided on Table II, which shows edge node 5 having the minimal CPU usage rate (or 0.07) among the list of encountered edge resources.

In addition, Edgify identifies $e_6$, followed by $e_3$ then $e_2$ in the list of ranked resources based on the Edgify scores. These results align with the data shown in Table II where $e_6$, $e_3$ and $e_2$ have CPU usage rates of 0.13, 0.15 and 0.18, respectively. Unlike with the case of the memory-centric, for CPU-centric, Edgify attempts to find resources having lower CPU usage rates.

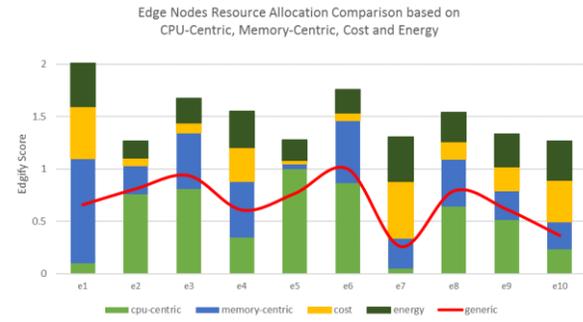

Fig. 3. Results from CPU-, Memory-Centric, Cost and Energy Task Allocation Experiments

**Conclusion**

In this paper, we presented Edgify, a framework for efficiently allocating edge resources within Internet of Medical Things (IoMT) environments. By dynamically optimizing the resources at the edge of IoMT networks, it is then possible to solve the challenges associated with the having high latency rates when offloading tasks to the cloud. We evaluated our proposed optimization and resource allocation approach using the GWA-T-12 Bitbrains dataset. Results demonstrate the effectiveness of employing multiple-criteria decision analysis methods such as TOPSIS when allocating edge-based resources for executing tasks. For future work, we plan to extend our Edgify framework to support mobility and considering time factors associated with the offloadability of tasks as part of the optimization process.